# Energy Balance in Cell Phone Radiofrequency Radiation Exposed Mice and Rats

Bernard J. Feldman

Department of Physics and Astronomy

University of Missouri-St. Louis

St. Louis, MO  63121

Feldmanb@umsl.edu

314-516-6805

Abstract

The National Toxicology Program exposed mice and rats to cell phone radiofrequency radiation.  They observed cancers in male rats but not in male mice and different increases in body temperature between mice and rats using identical frequency and intensity exposures.  Using the Faraday's Law model, thermoregulation considerations and an energy balance equation, we estimate that the increase in the temperature of neural circuits of mice exposed to radiofrequency radiation in the National Toxicology Program is only 0.28 that of rats.  This decrease in neural circuit temperatures is consistent with the absence of cancer in mice.  This calculation also qualitatively explains the different increases in body temperature in mice and rats under different radiofrequency intensities and frequencies.

The National Toxicology Program (NTP) issued its final report in 2018 on cancer in Sprague Dawley rats and B6C3F1 mice exposed to cell phone radiofrequency radiation.[1]  The experiments involved exposing mice to between 0 and 10 W/kg of 1,900 MHz radiation and rats between 0 and 6 W/kg of 900 MHz radiation.  The head of the study, Dr. John Bucher, was asked why different radio frequencies were used for mice and rats.  He replied, "900 MHz frequency was absorbed more readily by rats and the 1,900 MHz frequency was absorbed more readily by mice."[2]  The NTP also reported two related results:  first, the radiation exposure of all animals never caused a greater than $1°C$ increase in body temperature; second, cancers were found in male rats but not in male mice.[1]

The starting point in the discussion of these NTP results is the model proposed by this author to explain how non-ionizing radiofrequency radiation can cause cancer. The basic idea is neural closed loop circuits absorb radiofrequency radiation due to Faraday's Law.[3,4] Assume a radiofrequency magnetic field $B = B_o \cos(\omega t)$, where $B_o$ is the amplitude of the magnetic field and $\omega/2\pi$ = frequency of the radiation. If there is a circular neural closed loop with radius r, from Faraday's Law, the induced voltage around that closed loop is $V = \omega B_o \sin(\omega t)\pi r^2$. Then by Ohm's Law, the heat generated per second in that closed loop $P = V^2/R$, where R is the resistance in the loop. The heat generated per second per unit length of the neural circuit, P/r, is proportional to $B_o^2 \omega^2 r^2$. (The radii of the neurons in rats and mice are approximately the same, so the neural cross-sectional area is not a function of r and not included in the calculation.) It is this proportionality that will go into the energy balance equation to determine the increase in the neural circuit's temperature.

   The second part of the model is how this very localized heat in the neural circuit can cause cancer. The model proposes that in these neural circuits there is present both carcinogenic radicals, like oxidants, and antioxidants—molecules that bind to and eliminate carcinogenic radicals. The heat generated in these neural circuits raises the temperature of these circuits which either increases the concentration of carcinogenic radicals and/or decreases the concentration of antioxidants, thus increasing the chances that carcinogenic radicals will cause cancer near or in these neural circuits.

    The new idea in this manuscript is that it is the energy balance between heat absorption and heat removal that determines the final neural circuit temperature. Faraday's Law model gives the relationship between heat absorption and frequency, intensity (intensity is proportional to $B_o^2$) and size. A discussion of thermoregulation in nice and rats is needed to find the relationship between heat removal and size, which the author discussed in a previous manuscript.[5] The bottom line of that manuscript is that mice are far more efficient than rats in heat removal. This is not only because of theirs smaller size (smaller animals have a higher ratio of surface area of lungs, skin, and ears to their volume), but also mice have a higher heart rates than rats, thus moving heat

from hot neural circuits to cooler lungs at a faster rate.   But there are other factors at work.   For example, rats have a hairless tail that removes about 17% of the heat from their bodies, but mice do not.   As an aside, the evolution of hairless rat tails is evidence that ancient rats had problems with heat removal as they grew in size over millions of years.   It was also suggested that this same dynamic was partly at work in the evolution of human hairlessness.[5]

Using the energy balance equation that equates the heat absorption rate P/r to the heat removal rate, we get a relationship between temperature increase of the neural circuit, intensity, frequency and size.   If we assume that the heat removal rate is proportional to the increase in temperature (a heat conduction model) and inversely proportional to size (the ratio of surface area to volume), then the temperature of the neural circuit is proportional $B_o^2\omega^2 r^3$.

Let us now apply this result to the Sprague Dawley rats and B6C3F1 mice.  B6C3F1 mice are about one third the size of Sprague Dawley rats.  The mice were exposed to a higher frequency radiation (1900 MHz vs 900 MHz) and higher intensity (10 W/kg vs 6 W/kg) than the rats. Using the energy balance result, we calculate that the temperature increase in the neural circuits of mice is about 0.28 the temperature increase in the neural circuits of rats.   This is consistent with the lack of any observed cancers in male mice compared to male rats.

This result can also qualitatively explain why a higher frequency and intensity of radiofrequency radiation was needed to increase the body temperature of mice by 1 ºC compared to that of rats.   Increasing the frequency and intensity increased that absorption by those neural circuits and increasing the intensity also increased the absorption by the rest of the mice's body.

In conclusion, the Faraday's Law model can explain many of the results of the NTP study, including the differing cancer rates between rats and mice and differing increases in body temperatures of rats and mice.   This model also predicts that the cancer rates should be roughly linear in intensity (proportional to $B_o^2$) and this is also observed.[1,6]   Cancers detected in male rats were gliomas in brains and schwannomas in heart neural cell sheaths.[1]  No cancers were found in large organs not surrounded by neurons like lungs, pancreas or liver or in small organs like thyroid, prostate or kidneys.  This is consistent with the Faraday's Law model where the heating is most pronounced near nerves in or around large organs.